\documentclass[12pt, preprint]{aastex}

\shorttitle{Deep Meridional Circulation}
\shortauthors{Braun and Birch}

\begin{document}

\title{Prospects for the Detection of the Deep Solar Meridional Circulation}

\author{D.~C.\ Braun, A.~C.\ Birch}
\affil{NorthWest Research Associates, CoRA Division, 3380 Mitchell Lane, 
Boulder, CO 80301, USA}
\email{dbraun@cora.nwra.com aaronb@cora.nwra.com}

\begin{abstract}
We perform helioseismic holography to assess the noise
in p-mode travel-time shifts which would form the basis of
inferences of large-scale flows throughout the
solar convection zone.
We also derive the expected travel times from a parameterized
return (equatorward) flow component of the meridional circulation 
at the base of the convection zone 
from forward models under the assumption of 
the ray and Born approximations. 
From estimates of the signal-to-noise ratio 
for measurements focused near the base of
the convection zone, we conclude that the helioseismic detection of 
the deep meridional flow including the return component
may not be possible using data spanning an interval less
than a solar cycle.

\end{abstract}

\keywords{Sun: helioseismology, interior}

\section{Introduction}

Among all known large-scale flows in the Sun, the
meridional circulation has particular significance
because of its role in the transport of angular momentum and
magnetic flux across a wide range of latitudes within the
convection zone. Consequently, it is a significant component of
models of the dynamics
of rotating stellar convection zones, dynamos, and the solar cycle
\citep{Glatzmaier1982, Choudhuri1995, Dikpati2001, Wang1991, 
Wang2002, Hathaway2003, Hathaway2004, Dikpati2006, Dikpati2007}.

Measurements of the surface manifestation of meridional
circulation have typically indicated poleward flows
between 10 and 20 m s$^{-1}$ \citep[e.g.][]{Hathaway1996}.
Although frequencies of global p modes are insensitive (to
first order) to the meridional circulation, the flows have been
detected with a variety of local seismic methods 
\citep[e.g.][]{Giles1997, Giles1998, Braun1998, Gonzalez1999,
Giles2000, Haber2002, Hughes2003, Zhao2004,
Chou2005, Gonzalez2006, Svanda2007, Mitra-Kraev2007, Gonzalez2008}.
Many of these studies have focused their attention on
the meridional circulation near the surface (e.g.\ within a few
tens of Mm below the surface), and only a few attempts
have been made to deduce the properties of the deeper
components. Among the most comprehensive analyses is the work
of \citet{Giles2000} which is based on models of time-distance measurements
using over two years of Michelson Doppler Imager 
\citep[MDI;][]{Scherrer1995} 
Dopplergrams. These models included two general
solutions for the meridional circulation as a function of depth
and latitude: the first (hereafter ``Giles' model A'') without any
constraint on mass conservations, and the second (Giles' model B)
with an imposed mass conservation. Only Giles' model B exhibited a return
flow while the other showed exclusively poleward flows throughout
the convection zone. Each model was consistent with the travel-time
measurements within their range of errors \citep{Giles2000}.

In the frequency--wavenumber range of p modes 
propagating through the bottom half of the convection zone,
the random noise present in most current helioseismic measurements
is dominated by realization noise caused by stochastic excitation of the 
p modes near the solar surface. For the exploration of large-scale
flows such as meridional circulation this can be reduced by observing
more of the Sun (e.g.\ the far side of the Sun which is not currently 
accessible to helioseismic instruments) or by employing datasets 
with longer temporal duration.
With over a decade of helioseismic observations from both 
the Global Oscillations Network
Group (GONG; \url{http:/gong.nso.edu/data}) and MDI (\url{http:/soi.stanford.edu/data}) now available it is worthwhile to revisit the issue of 
the deep meridional circulation.
In this paper, we explore the prospects
for helioseismic detection of the return component of the meridional 
circulation in the deep
solar convection zone by applying helioseismic holography
to MDI observations to assess the random noise
in travel-time shifts which would form the basis for the
inference of large-scale flows.  Our analysis and resulting 
noise estimates are described in 
\S~\ref{noise}. We estimate the expected signal from
a plausible return component of meridional circulation using forward modeling procedures 
described in \S~\ref{models}.  
This is followed in \S~\ref{discussion} by a discussion of
the implications of these results.

\section{Noise Assessment} \label{noise}

Helioseismic holography (hereafter HH) is a method which
computationally extrapolates the surface acoustic field
into the solar interior \citep{Lindsey1997, Lindsey2000}
in order to estimate the amplitudes of the waves propagating into
and out of a focus point at a chosen depth and position in the solar
interior. The magnitudes and phases of these amplitudes, 
called the ingression and egression, are used to detect
flows and other perturbations to the waves.
The method employed for horizontal flow diagnostics is based on the
egressions and ingressions computed in the {\it lateral
vantage} employing pupils spanning 4 quadrants
extending in the east, west, north and south directions
from the focus \citep{Braun2004, Lindsey2004, Braun2007}.
In the lateral vantage, the p modes sampled by the pupil
propagate through the focal point in directions inclined up
to $\pm 45^\circ$ from the direction parallel to the
surface \citep[see Figure 3 of][]{Braun2007}.
A difference in the travel times between
waves traveling from one pupil to its opposite and
waves traveling in the reverse direction
is produced by flows along the path of the waves. 
In particular, the travel time differences, 
$\delta{\tau}_{\rm ew}$ and $\delta{\tau}_{\rm ns}$
derived from the east--west
and north--south quadrant pairs, respectively,
provide the HH signatures sensitive to the two components of
the horizontal flow.
The sign of the travel-time difference is such that a 
northward velocity component will produce a negative
value of $\delta{\tau}_{\rm ns}$.
The lateral-vantage geometry samples  
more than 70\% of the wave modes which pass through
the focus. The remaining waves, propagating more vertically than
the waves appearing in the pupil, are substantially less sensitive to 
horizontal flows near the focus.
Table~\ref{tbl-1} lists the focus depths and the pupil 
radii used in this study. The pupil radii are
defined from ray theory. The range
of (spherical-harmonic) mode degrees ($\ell$) at
4 mHz, selected
by each pupil, is also listed in the table. The lower
$\ell$ value denotes the modes propagating at
$\pm 45^\circ$ 
from the horizontal direction which propagate through the
focus and reach the surface at either the inner or
outer pupil radius. The highest $\ell$ value listed
indicates modes propagating horizontally through the focus. 
The focus depths extend down to the base of the convection zone. 
However, the analysis is conceptually similar to previous near-surface
measurements \citep{Braun2007}.

Three weeks of full disk Dopplergrams with one minute
cadence, obtained from MDI
were used in this study.
The data set spans the interval from 1996 June 25 to July 16,
and coincides with a period of very low magnetic activity 
on the Sun. Smaller spans of data at other epochs (2002 March
and 2003 October) were also
examined. Travel-time maps made at all three epochs
exhibit similar noise characteristics, and 
we show here only the results using the 1996 data.
The following steps summarize the general data reduction:
1) a projection of each 24 hr segment of full-disk data onto 
nine Postel projections (each extending $180^\circ \times 180^\circ$) 
centered on grid points separated by $40^\circ$ in 
heliographic latitude and 
central meridian distance referenced to midday
and rotating with the Carrington rate,
2) temporal detrending by subtraction of a linear
fit to each pixel signal in time, 3) removal of poor
quality images, identified by a five-$\sigma$ deviation of
any pixel from the linear trend , 4) Fourier transform of
the data in time,
5) extraction of the frequency
bandpass spanning 2.5 to 5.5 mHz, 6) computation of Green's
functions over the appropriate pupil, 7) computation
of ingression and egression amplitudes by a 2D convolution
of the data with the Green's functions, 
8) computation of the travel-time difference maps, and
9) extraction and remapping of the central $40^\circ \times 40^\circ$ portion of
each region to form mosaics in heliographic coordinates
spanning $120^\circ \times 120^\circ$.
The Green's functions (step 6) were computed using 
the eikonal approximation \citep[][]{Lindsey1997, Lindsey2000} in
spherical coordinates. The large size of the Postel's projections
is dictated by the large pupil required for the deepest focii
in Table~\ref{tbl-1}.
The 2D convolution (step 7) is a time-saving convenience,
appropriate for the type of preliminary noise estimates we are 
interested in, but distorts the resulting travel-time difference maps. 
This occurs because of a geometrical 
mismatch between the fixed annular pupil assumed for the 
convolution operation and the correct 
pupil whose shape varies with position in the Postel projection.
This deviation worsens as the horizontal position of the focus is moved
away from the central (tangent) point of the Postel projection,
and the effects of this can therefore be constrained 
by combining maps made using multiple locations of
the tangent point.  The consequences of this distortion
for the results presented here are discussed below. 

\begin{figure}[htbp]
\epsscale{1.0}
\plotone{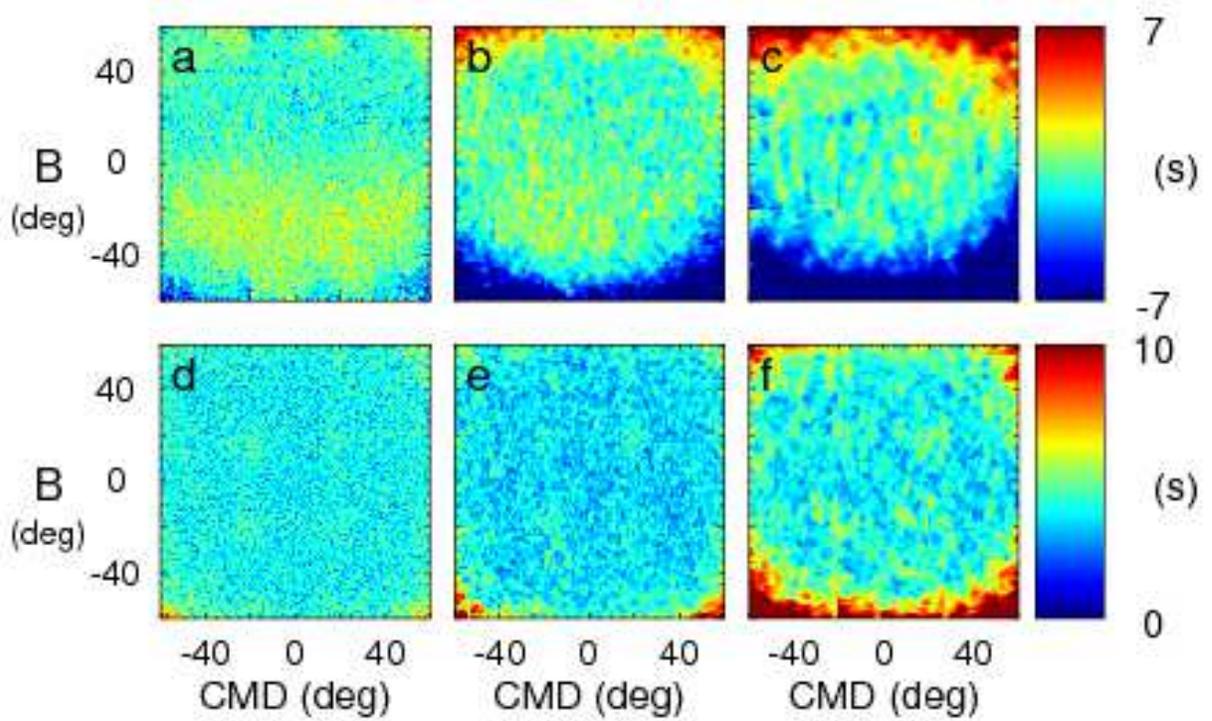}
\caption{Maps of the mean (panels a-c) and standard deviation 
(panels d-f) of north-south travel-time differences for twenty
consecutive 24-hr 
measurements. From left to right the focus depths of
the measurements are 30, 100, and 200 Mm respectively.
}
\label{maps}
\end{figure}

Figure~\ref{maps} shows selected maps of the mean 
$\delta{\tau}_{\rm ns}$ and standard deviation $\sigma_0$ of the north-south 
travel-time difference
over twenty consecutive (24-hr duration)
sets of measurements. Each map covers an area
of $120^\circ$ in central meridian distance (CMD) and
heliographic latitude (B).
The maps for one day of data (June 27) were
not included in further analysis due to an anomalously high amount of
poor images. Near the surface (e.g.\ Figure~\ref{maps}a), 
the meridional flow produces a distinct 
negative (positive) travel-time difference in the north (south) hemisphere.
As the focus depth increases, this signature becomes less visible.
A distinct pattern near the poles is also evident and increases
significantly with greater focus depth. This pattern is opposite
in sign of the meridional signature and is clearly an artifact
centered on the position of disk center (about $+3^\circ$) 
as observed by MDI.

Remarkably, the maps of the standard deviation (Figs~\ref{maps}d -
\ref{maps}f) indicate that, apart from the vicinity of the solar limb, 
the noise for a single travel-time measurement is fairly
constant ($\sigma_0 \approx 4 \rm{sec}$) with focus depth. 
There is a granularity in these
maps which becomes courser with depth and is related to
the increase in the horizontal wavelength of the modes
used to make the measurements. An increase in the
standard deviation near the solar limb is evident.
In addition, there is a noticeable 
excess of $\sigma_0$ near positions 
(CMD, B) = $(\pm 20^\circ, \pm 20^\circ)$ which are
the corners of the individual subregions of 
the mosaic. The noise in these corners is about 15\%
above the values near the Postel tangent points for all
focus depths used here. This feature likely
results from the use of the 2D convolution of Postel's
projections described earlier.  

Due to signal-to-noise issues we assume any current or future attempt to deduce 
properties of the deep meridional circulation will
make use of longitudinal averaging and very likely also involve at least
modest smoothing in latitude.  Figure~\ref{noise_vs_depth}
shows the standard deviation ($\sigma_{\rm a}$) of averages of $\delta \tau_{\rm ns}$ 
over strips spanning $120^\circ$ in CMD and $15^\circ$
in B. Unlike the standard deviation ($\sigma_0$) corresponding to specific 
horizontal focus positions which show little or no variation with
focus depth, the standard deviation of the mean ($\sigma_{\rm a}$)
increases with depth at all latitudes. This
is a consequence of having fewer independent measurements within a fixed
area as the depth increases and is analogous to the common problem in global
helioseismology of having fewer modes in which to deduce either
structural perturbations or flows at greater depths within the Sun.
There is also an increase in noise for measurements at high latitudes,
consistent with the maps of $\sigma_0$ (Figure~\ref{maps}).
Significantly, the values of $\sigma_{\rm a}$ for the $15 - 30^\circ$
strips are very close to the results for the $0 - 15^\circ$ strip. 
This offers some
assurance that the contribution of noise due to the 2D convolution 
(which should preferentially influence the 
$15 - 30^\circ$ strip)
does not add substantially to the results shown in 
Figure~\ref{noise_vs_depth}.

\section{Forward Models} \label{models}
 
We use both the Born and ray approximations to estimate the travel-time 
shifts that would be caused by a return flow near the base of the convection
zone. 
For a discussion of the ray approximation see \citet{Giles2000}.
We used the numerical approach of \citet{JCD89} for computing 
ray paths in spherical geometry.

For this letter we also make rough estimates of the 
sensitivity of HH travel times
to weak, steady, and horizontally uniform flows by approximating
the convection zone as a plane-parallel layer.
The functions which describe the linear 
sensitivity of the power spectrum to a horizontally uniform flow
can be computed using a generalization of the
Born-approximation based approach of \citet{Gizon2002} and \citet{Birch2007}.  
We used the normal-mode summation Green's functions from \citet{Birch2004}, 
though with the eigenfunctions for a spherical Sun in place of those for a 
plane-parallel version of model S.  
We used the source model of \cite{Birch2004}.
Changes in the the power spectrum may easily be related to changes in the
ingression-egression correlation through the expression for
the expectation value of the correlation. The result is a set of
sensitivity kernels which relate the correlations (and thus the
travel-time shifts) with the flows.

Using the sensitivity functions we estimate the travel-time shifts caused 
by deep return flows of the form $v(z) = A \cos\{\pi(r-r_{\rm c})/\Delta r\}$ 
for $r_{\rm c}<r<r_{\rm c}+\Delta r$ and $v(z)=0$ otherwise. 
Here $r_{\rm c}=496$ Mm is the radius of the base of the 
convection zone, $\Delta r$ is the thickness of the return flow, and $A$ 
is the maximum amplitude of the flow (Giles' model B can be roughly
approximated with $A \approx 3$~m s$^{-1}$ and  $\Delta r \approx 60$ Mm).
Figure~\ref{signal} shows travel-time differences, for two
focus depths, as a function of $\Delta r$, predicted from the two methods.
It is noteworthy that the times computed under the Born approximation
can differ substantially from those than predicted by the ray approximation. 
Much of the sensitivity in the Born approximation lies below the lower
turning point of the corresponding ray.
In addition, for a sufficiently thick return flow the Born travel-time
shifts are greater for the shallower
focus depth than for the deeper focus depth.
The deeper measurements use waves of higher phase speed, which undergo a
smaller phase shift in a horizontally uniform flow.

\section{Discussion}            \label{discussion}

\begin{figure}[htbp]
\epsscale{1.}
\plotone{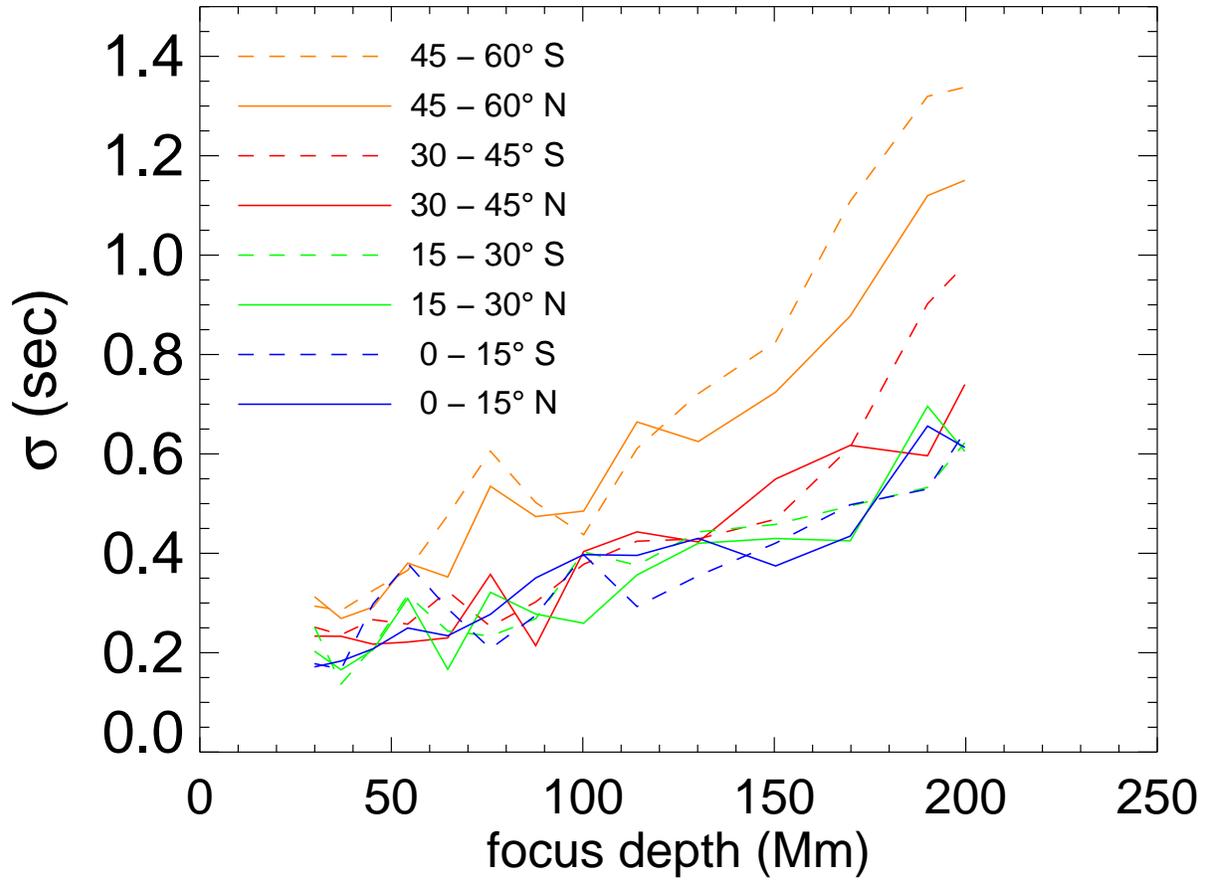}
\caption{
The standard deviation (over twenty consecutive 24-hr measurements) 
of the mean north-south travel-time difference 
averaged over a strip
of the Sun spanning 15 degrees in latitude and 120 degrees in central
meridian distance.
Different colors indicate different latitudes of the center of the strip,
while solid (dashed) lines indicate the southern (northern) hemisphere.
In general, the mean-standard-deviation increases with the depth of
the focus, and also increases at high latitudes.
}
\label{noise_vs_depth}
\end{figure}

\begin{figure}[htbp]
\epsscale{1.}
\plotone{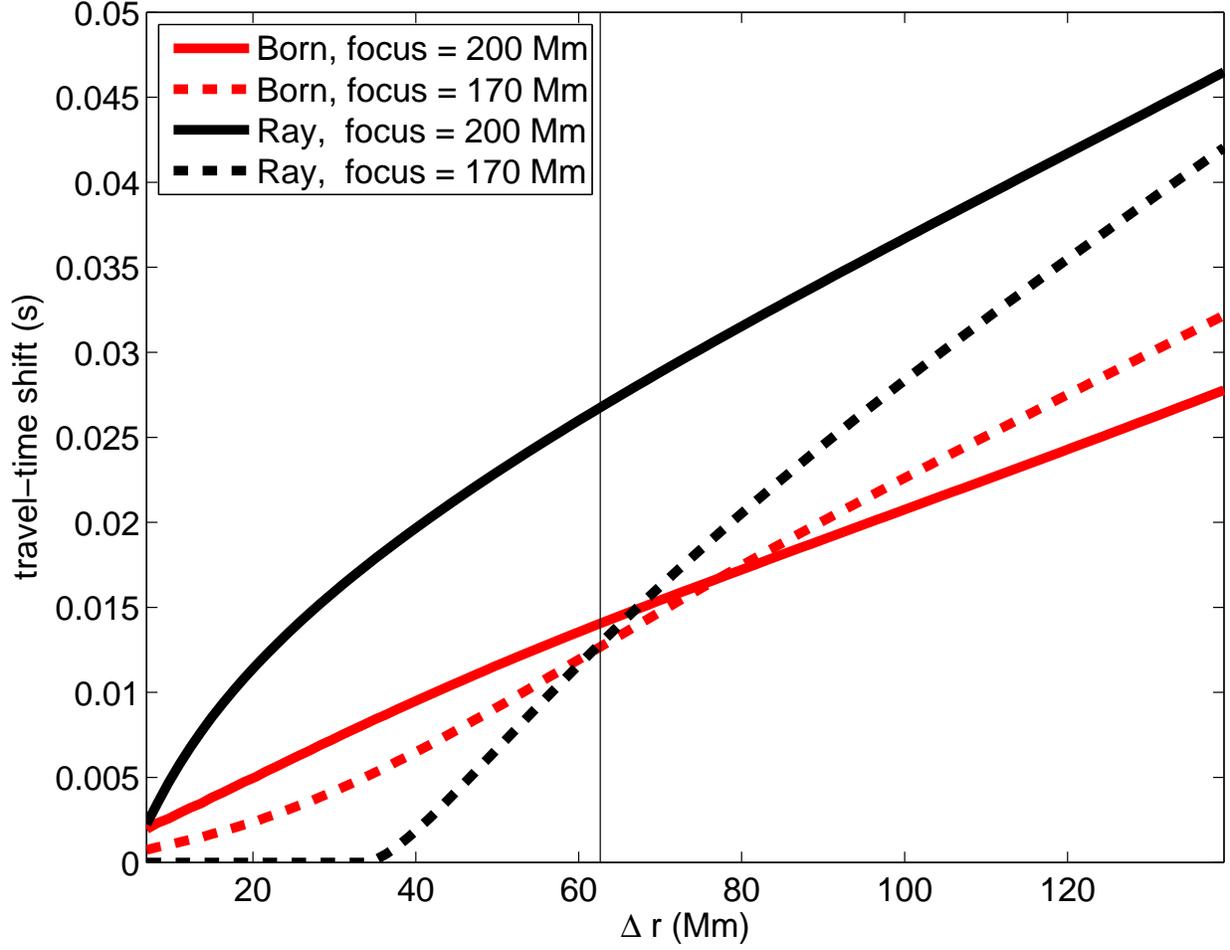}
\caption{
Expected north--south travel-time difference
as functions of the width of a 
hypothesized meridional return flow at the base of
the convection zone with a peak value of 3 m s$^{-1}$(see text). 
The flow is set to zero in the radiative zone.
The red (black) lines show the results of a Born (ray) 
approximation calculation, and the 
the solid (dashed) lines show the results for focus depths of 
200 (170) Mm below the surface.
The vertical line indicates the width which roughly corresponds to
Giles' model B. 
}
\label{signal}
\end{figure}

To estimate the amount of data needed for a detection of 
the return flow, we conservatively assume that
a successful detection requires a signal-to-noise ratio
(SNR) of three for travel-time measurements near the
base of the convection zone. Thus for the types of return flows
shown in Figure~\ref{noise_vs_depth}, we require a measurement
precision on the order of .01 seconds.  
Given a random noise of 0.6 seconds for a single day for
a HH measurement over a $15^\circ$ strip (Figure~\ref{noise_vs_depth}),
it is apparent that at least 12 years of uninterrupted data
is needed for a detection of a mean return flow with characteristics
similar to Giles' model B using the results of the ray approximation
(i.e.\ a travel-time shift of 0.027 seconds), while the
more confined flows to the left of the vertical line would require on the order of 
hundreds of years. These estimates are derived assuming the
error in the mean flows decreases with the square root of the
length of the time series \citep{Gizon2004}.
Combining data from both hemispheres and sacrificing some
latitudinal resolution (e.g.\ to $30^\circ$) it should be 
possible to (roughly) halve the required duration.
We therefore tentatively conclude that a 3$\sigma$ detection
of a return flow of a magnitude similar or smaller than
Giles' model B is possible only with an amount of data
comparable or greater than a solar cycle.

The additional
noise contributions or systematic errors due to analysis
artifacts or details of the modeling
procedures are not considered here, so that the 
results represent a ``best-case'' scenario.
We anticipate that high-quality measurements spanning a range of 
depths will actually
be needed to construct a model of the flow. 
Our emphasis on the SNR of the individual measurements
is based on the relative ``completeness''
of lateral-vantage HH in that a single ingression--egression
correlation  efficiently samples
and combines most of the waves propagating through a particular
depth. We therefore assume that the uncertainty in the inferred
flow at the base of the convection zone will be dominated by
the random noise contribution to measurements focused at that position.

It is worthwhile to compare our results with uncertainties in
other helioseismic measurements of flows near the base of the
convection zone. From inversions of global-mode frequency splittings,
\cite{Howe2000} derive 1$\sigma$ errors of inverted rotation
rates at a depth of 195 Mm of about 1 nHz (corresponding to 4.4 m s$^{-1}$) 
for 72-day sets of MDI observations with an averaging kernel 
approximately 50 Mm 
wide (FWHM) in depth and about $15^\circ$ wide in latitude.
A 3$\sigma$ detection of an average 1.5 m s$^{-1}$ zonal flow over this range
in depth and latitude would therefore require about 15 years of 
frequency-splitting measurements which agrees very well with the estimate presented here.

Although direct measurements of temporal variability of the
return component over timescales equal to or less than a solar
cycle appear unlikely, we note that some inferences about
variability could be made using shallower flow measurements
and assuming mass conservation, as did \citet{Giles2000}.
Regardless of possible temporal variability, 
it is likely well worth the effort to
carry out analyses and modeling of existing decade-length
helioseismic observations to either detect or constrain
the mean return component. However, we note that there are
a considerable number of potential qualifications to the rather
simple estimates derived here. Some of these involve issues of the resolution
of the models (e.g.\ the ability to infer
the existence of multiple meridional cells or sharp gradients 
of flow in latitude  or depth).
In addition, we expect considerable challenges to the probing of
any flows in the deep convection zone raised by the need to
understand and remove possible artifacts and systematic effects.
Some of these effects are visible as systematic differences 
between inferred flows using separate but contemporary datasets
\citep{Gonzalez2006}. These differences are often
of the order of a few m s$^{-1}$ which, while troublesome
for probing even the near-surface layers, are clearly
disastrous for the unambiguous identification of a return 
flow of comparable or smaller magnitude.
The critical need for multiple sources of long-duration helioseismic 
observations combined with careful artifact identification and 
correction procedures
in reducing both systematic effects and random
noise should be clear.

\acknowledgments
We appreciate useful discussions with M. Woodard.
DCB and ACB are supported by funding through NASA contract NNH05CC76C and
NSF grant AST-0406225, and a subcontract through the 
NASA sponsored HMI project at Stanford University awarded to NWRA.

\begin{deluxetable}{ccccc}
\tablecolumns{3}
\tablewidth{0pc}
\tablecaption{Pupil size and mode degrees\label{tbl-1}}
\tablehead{
\colhead{Depth} & \colhead{Pupil radii} & \colhead{$\ell$ @ 4mHz}\\ 
\colhead{(Mm)} & \colhead{(degrees)} & \colhead{ }\\ 
}
\startdata
30 &  1.3 - 10.5 & 166 - 235 \\
37 &  1.7 - 12.9 & 145 - 205 \\
45 &  2.0 - 15.4 & 128 - 180 \\
54 &  2.4 - 18.2 & 113 - 159 \\
64 &  2.9 - 21.1 & 100 - 142 \\
76 &  3.4 - 24.1 & 89 - 127 \\
88 &  4.0 - 26.5 & 81 - 114 \\
100 &  4.5 - 29.2 & 73 - 104 \\
114 &  5.2 - 32.0 & 66 - 93 \\
130 &  5.9 - 35.2 & 59 - 84 \\
150 &  6.8 - 41.4 & 52 - 74 \\
170 &  7.7 - 48.1 & 46 - 66 \\
190 &  8.6 - 54.9 & 41 - 58 \\
200 &  9.1 - 57.1 & 39 - 55 \\
\enddata
\end{deluxetable}

\end{document}